\documentclass[conference]{IEEEtran}
\IEEEoverridecommandlockouts

\usepackage{cite}
\usepackage{amsmath,amssymb,amsfonts}
\usepackage{algorithmic}
\usepackage{graphicx}
\usepackage{textcomp}
\usepackage{xcolor}
\usepackage{siunitx}
\def\BibTeX{{\rm B\kern-.05em{\sc i\kern-.025em b}\kern-.08em
    T\kern-.1667em\lower.7ex\hbox{E}\kern-.125emX}}
\begin{document}

\title{Optimal Duration of Reserve Capacity Ancillary Services for Distributed Energy Resources
}

\author{\IEEEauthorblockN{Lorenzo Zapparoli}
\IEEEauthorblockA{\textit{Department of Mechanical and} \\
                    \textit{Process Engineering} \\
\textit{ETH Zurich}\\
Zurich, Switzerland \\
lzapparoli@ethz.ch}
\and
\IEEEauthorblockN{Blazhe Gjorgiev}
\IEEEauthorblockA{\textit{Department of Mechanical and} \\
                    \textit{Process Engineering} \\
\textit{ETH Zurich}\\
Zurich, Switzerland \\
gblazhe@ethz.ch}
\and
\IEEEauthorblockN{Giovanni Sansavini}
\IEEEauthorblockA{\textit{Department of Mechanical and} \\
                    \textit{Process Engineering} \\
\textit{ETH Zurich}\\
Zurich, Switzerland \\
sansavig@ethz.ch}
}

\maketitle

\begin{abstract}
The increasing integration of distributed energy resources (DERs) into power systems presents opportunities and challenges for ancillary services (AS) provision.
Technical requirements of existing AS (i.e., duration, reliability, ramp rate, and lead time) have been designed for traditional generating units, making their provision by DER aggregates particularly challenging. 
This paper proposes a method to design the duration of reserve capacity AS products considering the operational constraints of DERs and the temporal dynamics of system imbalances. The optimal product duration is determined by maximizing product availability and aligning the supply profile with the system's balancing needs.
We apply the methodology to a realistic Swiss low-voltage network with a diverse DER portfolio. The results reveal that (i) shorter product durations maximize average availability and (ii) long product durations improve the alignment with system balancing needs. This paper offers valuable insights for system operators to design AS products tailored for DER participation.
\end{abstract}

\begin{IEEEkeywords}
ancillary services, virtual power plant, distributed energy resources, electricity market.
\end{IEEEkeywords}

\section{Introduction}
\label{sec:intro}
The integration of renewable energy sources (RES) and distributed energy resources (DERs) is radically transforming the power system, introducing significant challenges. Primarily, renewable energy generation's inherent variability causes increased demand for ancillary services (AS) to guarantee grid's reliability~\cite{xie2024, chicco2020}. One solution to this challenge is integrating DERs into AS markets through virtual power plants (VPPs)~\cite{pudjianto2007}, aggregating numerous small-scale DERs to efficiently trade energy and provide grid support services as a single, reliable resource. However, existing AS have been designed to match the characteristics of traditional generating units, namely thermal and hydro. These sources are programmable and highly reliable, while DERs are subject to strong uncertainties due to weather conditions and human behavior, and their availability is strongly time-dependent. This mismatch makes the provision of AS by DER aggregates challenging. This issue is particularly relevant for reserve capacity AS, which are traded with a large lead time, have strict reliability, and long duration requirements \cite{swissgrid2022_prequalification}. Transmission system operators (TSOs) are looking for new AS designs that maximize the potential quantity made available by DERs while meeting the balancing needs. For this reason, AS markets are undergoing significant transformations~\cite{RANCILIO2022111850}. Key trends driving these changes include decreasing the time definition of products to allow for shorter service durations, thereby allowing more flexible resource management. Similarly, shortening the interval between market closure, and product activation and relaxing reliability requirements reduces the uncertainty players have to cope with in the bidding process. Furthermore, lowering the minimum bid size enables smaller-scale players to participate.

In this context, the Danish TSO, Energinet, has introduced a revised set of requirements for reserve capacity ancillary services to facilitate the integration of non-programmable resources~\cite{energinet2023prequalification}. A notable modification is the relaxation of the reliability requirement, which has been reduced from \SI{99}{}–\SI{99.9}{\%} to \SI{90}{\%}. This adjustment permits a reserve shortfall probability of up to \SI{10}{\%}, thereby incentivizing the participation of stochastic flexible resources. The authors in~\cite{gade2024leveragingp90requirementflexible} refer to this new prequalification criterion as the P90 requirement and analyze its implications using an electric vehicle aggregator as a case study. They formulate a distributionally robust joint chance-constrained program to optimize reserve capacity bids, demonstrating that the P90 requirement enhances the aggregate's ability to provide reserve capacity under uncertainty. The results remark the need for ancillary services that account for the characteristics of stochastic resources. The design should foster reserve availability but also fulfill the system's balancing needs. Furthermore, future research directions are suggested, including the application of these insights to heterogeneous resource portfolios and the exploration of additional requirements.

In this paper, we investigate another reserve capacity product's requirement: the duration. It represents the length of the time windows in which the product is offered. It is widely accepted that reducing the duration of ancillary service products can increase flexibility provision by DERs~\cite{RANCILIO2022111850}. However, to the best of the authors' knowledge, no studies in the literature have specifically investigated power reserve AS duration as a design variable. While shorter product durations may increase the overall quantities of reserves available, they can also result in an undesirable temporal shift in reserve availability. This shift may cause reserves to be available during periods of lower system demand while being insufficient during peak demand periods. The time-dependent characteristics of DERs exacerbate this issue. This necessitates ancillary service designs that not only align with DER capabilities but also satisfy the balancing requirements of TSOs. Additionally, most studies focus on a single type of DER technology, neglecting the potential benefits and challenges introduced by technological diversification within DER portfolios.

To address these gaps, this paper proposes a method to optimally design the duration requirement of reserve capacity products.
We apply the methodology to a VPP built on a realistic Swiss low-voltage network with a diversified DER portfolio. The considered VPP includes distributed generators (DGs), electric vehicles (EVs), battery energy storage systems (BESS), and heat pumps (HPs). The proposed method involves three main steps. First, historical imbalance data of the studied national power grid is processed to extract representative imbalance profiles. Second, the VPP reserve capacity product supply potential profiles are characterized as a function of the investigated parameter, in this case, product duration. This step involves modeling all DERs within the VPP and estimating the maximum product quantity they can provide under constraints related to DERs operations, distribution network limits, and AS product's ramp rate, reliability, direction, and duration. Forecast uncertainty is treated in the supply potential estimation problem through Monte Carlo simulation. Third, the optimal product duration is determined based on two objectives: maximizing the average product availability and aligning the imbalance and reserve supply profiles.

The contributions of this paper are:  
1) it proposes a novel method to design the optimal duration of a reserve capacity ancillary service product, optimizing availability and aligning it with TSO balancing requirements;
2) it assesses the impact of product duration on the provision capability profile of a technologically-diversified VPP;
3) it explores the trade-off between the two objectives and proposes an optimal duration for the DER-oriented upward and downward reserve capacity products for the Swiss case study.

The remainder of this paper is organized as follows. Section~\ref{sec:method} describes the main features of the proposed methodology. Section~\ref{sec:cstudy} defines the case study. Section~\ref{sec:results} provides results followed by a discussion thereof. Section~\ref{sec:conclusions} concludes the paper with remarks about limitations and future work.

\section{Method}
\label{sec:method}
The procurement of power reserve capacity AS is carried out by TSOs through tendering procedures that occur $t^{\text{lead}}$ before the delivery period $T^{\text{de}} = [t^{\text{de}}, t^{\text{de}} + t^{\text{p}}]$, where $t^{\text{de}}$ is the delivery start time and $t^{\text{p}}$ is the product duration. Reserve capacity products are further defined by their direction $d^{\text{p}}$, a minimum ramp rate requirement $r^{\text{p}}$, and a reliability requirement $R^{\text{p}}$~\cite{swissgrid2022_ASproducts}.

This section outlines a methodology to optimize the duration $t^{\text{p}}$ of AS reserve capacity products to enhance their availability from DERs while aligning with TSO balancing requirements. As introduced in Section \ref{sec:intro}, the approach consists of three steps: processing historical imbalance data to extract representative profiles, modeling the reserve capacity supply potential of a DER aggregate while accounting for operational and market constraints, and determining the optimal product duration by aligning DER supply potential with system imbalance profiles and maximizizing the availability.

The remainder of this section is organized as follows. Section~\ref{sec:method_imbalance_profiles} describes the characterization of the TSO balancing needs. Section~\ref{sec:method_reserve_supply} describes the model that estimates the reserve capacity product supply by the VPP parametrized to the product duration. Lastly, section~\ref{sec:method_product_design} presents the approach to determine the optimal product duration based on the derived imbalance and product supply profiles.

\subsection{Imbalance profiles characterization}
\label{sec:method_imbalance_profiles}

Power reserves are largely used to manage power imbalances. Therefore, the characterization of a representative imbalance profile provides insights into the time dependence of the reserve demand, enabling us to consider system needs in the AS design problem. Unfortunately, direct data on imbalance profiles are unavailable or sparsely reported for Switzerland and many other European countries. To address this limitation, a proxy for the power imbalance must be identified. Secondary reserve energy activation was selected as a viable solution. This choice stems from the observation that primary reserves are also activated for cross-border imbalances and tertiary reserves are often used for purposes beyond imbalance management, such as congestion resolution, and with significant delay to the imbalances themselves. Secondary reserve energy activation data for Switzerland and several European countries are available on the ENTSO-E Transparency Platform~\cite{ENTSOE_transparency}. We extract representative daily profiles that describe average temporal dynamics of imbalances, with a focus on capturing intra-day and seasonal variability. Specifically, four days are derived, one for each season. Each day is characterized by its hourly season-average upward and downward reserve activation profiles.  

The data processing involves the following steps. First, data cleaning on the secondary reserve activation database is performed, i.e., outliers are identified and removed, and missing entries are imputed. Second, hourly values were computed by summing the corresponding four 15-minute entries for each hour. Third, for each season $s \in \mathcal{S} = \{\mathrm{spring, summer, autumn, winter}\}$, for each direction $d^{\text{p}} \in \mathcal{D} = \{\mathrm{upward, downward}\}$, and for each hour of the day $h \in \mathcal{H} = \{1, \ldots, 24\}$ the average activation value is computed as:  
\begin{equation}  
P_{s,h}^{\text{act}, d^{\text{p}}} = \frac{1}{|\mathcal{Y}||\Delta_s|} \sum_{y \in \mathcal{Y}} \sum_{\delta \in \Delta_{s}} \sum_{t \in \mathcal{T}_{y,\delta,h}} P_{t}^{\text{act}, d^{\text{p}}}
\end{equation}  
where $P_{s,h}^{\text{act}, d^{\text{p}}}$ represents the average power activation for direction $d^{\text{p}}$, season $s$, and hour $h$. Here, $\mathcal{Y}$ is the set of the considered years, $\Delta_s$ is the set of days in season $s$, and $\mathcal{T}_{y,\delta,h}$ is the set of time steps of the reserve activation profile in year $y$, day $\delta$, and hour $h$. Lastly, $P_{t}^{\text{act}, d^{\text{p}}}$ is the reserve activation value in direction $d^{\text{p}}$ and time step $t$. This procedure allows to reduce the time series length and filter noise while preserving daily and seasonal features of the data.

\subsection{VPP Reserve capacity supply characterization}
\label{sec:method_reserve_supply}
This section details the methodology for evaluating the VPP's reserve capacity supply potential. It generates a family of supply profiles, parameterized by ancillary service product duration $t^{\text{p}}$, considering the operational constraints of DERs, network limitations, and product specifications. This is achieved by dividing each season-representative day into a set of non-overlapping time windows $\Psi$ of duration $t^{\text{p}}$. For each time window $\psi \in \Psi$, we determine the maximum reserve capacity $q_{\psi}^{\text{p}}$. This process is repeated for all representative days and each $t^{\text{p}} \in \Omega$, where $\Omega$ is the set of the considered product durations. This generates a family of reserve capacity availability profiles $P_{t^{\text{p}},s,h}^{\text{ava}, d^{\text{p}}}$ as a function of product duration $t^{\text{p}}$, direction $d^{\text{p}}$, season $s$, and hour $h$. 

Some input parameters, grouped in the set $\mathcal{P}^{\text{u}}$, are subject to forecast uncertainty because the power reserve is scheduled $t^{\text{lead}}$ ahead of delivery. This influences the maximum reserve capacity the VPP can provide due to the reliability requirement $R^{\text{p}}$. To address this challenge, we employ a Monte Carlo simulation with Latin hypercube sampling~\cite{dong_quantile_2017}. For each realization of the uncertain parameters $p \in \mathcal{P}^{\text{u}}$, a deterministic reserve maximization problem is solved, yielding a corresponding maximum reserve quantity $q^{\text{p},*}_{\psi}$. By repeating this process across a large number of realizations, the distribution of the maximum reserve quantity is characterized. The reserve capacity available for bidding, $q_{\psi}^{\text{p}}$, is then defined as the $\alpha^{\text{p}}$-quantile of this distribution, which guarantees the reliability requirement $R^{\text{p}}$ of the product, where $\alpha^{\text{p}} = 1 - R^{\text{p}}$.


The reserve maximization problem, given a realization of the uncertain input parameters, defines the corresponding maximum reserve quantity $q^{\text{p},*}_{\psi}$. It identifies the network's and DERs' operational state that maximizes the power reserve provision and is formulated as follows:

\begin{equation}
\begin{aligned}
& \underset{\boldsymbol{x}}{\text{max}} & & q^\mathrm{p} \\
& \text{s.t.} & & \text{Product-defining constraints} \\
& & & \text{Network constraints} \\
& & & \text{DER operational constraints}
\end{aligned}
\end{equation}
where $\boldsymbol{x}$ is the vector of decision variables representing the operational state of the network and the DERs. The objective of this optimization problem is to maximize the power reserve product quantity $q^{\text{p}}$, subject to three distinct sets of constraints. The first set, referred to as product-defining constraints, maps electrical quantities into the corresponding product quantity. The second set consists of the network constraints, which model the VPP distribution network and enforce voltage and flow limits. The third set comprises the DER constraints, which model the operational limits and the technical characteristics of the active resources part of the VPP. 

The VPP is simulated in three states to guarantee feasibility: the dispatching state (disp), which reflects the position acquired in the energy markets; the activated upward reserve state (up), which represents the scenario where the TSO fully utilizes the booked upward reserve; and the activated downward reserve state (down), which represents the scenario where the TSO fully utilizes the booked downward reserve. 
The product-defining constraints are:
\begin{subequations} \label{eq: product_duration_constraints}
\begin{align}
        q^{\text{p}} &\leq q^{\text{p}}_t  \quad & \\
        q^{\text{p}}_t &= P_{t,\text{up}}^{\text{PCC}} - P_{t,\text{disp}}^{\text{PCC}}  \quad &\forall t \in \mathcal{T}_{\psi} \\
        q^{\text{p}}_t &= P_{t,\text{disp}}^{\text{PCC}} - P_{t,\text{down}}^{\text{PCC}}  \quad &
\end{align}
\end{subequations}
where $\mathcal{T}_\psi$ is the set of time steps within time window $\psi$. The time resolution is set to 1 hour, matching the resolution of the representative imbalance profiles. These constraints ensure that the product quantity $q^\text{p}$ is available for each time step $t$ during the delivery period. The product quantity \(q^{\text{p}}\) is defined as the difference between the power exchanged at the point of common coupling $P_{t}^{\text{PCC}}$ in the dispatching state and the upward and/or downward activated reserve states.

The network is modeled in all states using the linear DistFlow equations~\cite{19266}. This formulation leverages the radial topology assumed for the VPP distribution network to provide a direct linearized version of the power flow equations. These constraints relate to the power injected by every DER to $P_{t,s}^{\text{PCC}}$ and model voltage and flow limits of the grid. 

Single DERs are modeled using mathematical constraints that represent their operations. Different sets of constraints are applied to the four technology classes: distributed generators, heat pumps, battery energy storage systems, and electric vehicles. For DGs, the real power output can be controlled from zero to the maximum power, computed as the product of the unit's nominal power and a capacity factor modeling the primary resource availability. Heat pumps are modeled to maintain comfortable indoor temperatures, controlling the heating power considering building properties (thermal resistance and capacitance) and ambient temperature. Electric vehicles are modeled based on charging events, considering vehicle-to-grid operations. BESS and EVs are modeled with fixed charging and discharging efficiencies. Temperature for HPs and SOC for BESS are required to be the same at the beginning and end of the observed period, allowing the resources to be available in the subsequent time steps. A minimum average charge rate is required for electric vehicles. DGs and BESS can exchange reactive power (limited by the inverter rating) and EVs and HPs can not. 

The problem is formulated as a mixed-integer linear program (MILP). The parameters considered subject to uncertainty are the generator capacity factors, ambient temperatures, EV presence parameters, and non-dispatchable load powers. Representative days are selected as the central day of each season. To reduce computational complexity, spring and autumn were modeled using a single season-representative day.
%
In conclusion, solving for all durations, directions, time windows, and seasons, we can characterize the reserve capacity availability profiles $P_{t^{\text{p}},s,h}^{\text{ava}, d^{\text{p}}}$ as a function of the duration requirement $t^{\text{p}}$.

\subsection{Product duration design}
\label{sec:method_product_design}
To assess the optimal duration of a product, we aim to achieve two distinct goals. The first goal is to maximize the availability of the reserve product. The second goal is to ensure that the reserve capacity supply profile matches the reserve demand one.

The product duration design is formulated as 
a multi-objective optimization problem with two objectives. The first objective is maximizing product availability, quantified as the average reserve capacity over all time steps:
\begin{equation}
\underset{t^\text{p}}{\text{max}} \quad \frac{1}{|\mathcal{H}||\mathcal{S}|} \sum_{s \in \mathcal{S}} \sum_{h \in \mathcal{H}} P_{t^{\text{p}},s,h}^{\text{ava}, d^{\text{p}}}
\end{equation}
The second objective aims to align the supply profile with the demand profile. This is achieved by minimizing the mean squared error between the supply and demand curves. The errors are computed after normalizing both curves to their maximum values:
\begin{equation}
\underset{t^\text{p}}{\text{max}} \quad - \frac{1}{|\mathcal{H}||\mathcal{S}|} \sum_{s \in \mathcal{S}} \sum_{h \in \mathcal{H}} \big( \tilde{P}_{t^{\text{p}},s,h}^{\text{ava}, d^{\text{p}}} - \tilde{P}_{s,h}^{\text{act}, d^{\text{p}}} \big)^2
\end{equation}
where the normalized supply and demand profiles are defined as:
\begin{equation}
\tilde{P}_{t^{\text{p}},s,h}^{\text{ava}, d^{\text{p}}} = \frac{P_{t^{\text{p}},s,h}^{\text{ava}, d^{\text{p}}}}{\max(P_{t^{\text{p}},s,h}^{\text{ava}, d^{\text{p}}})}
\end{equation}
\begin{equation}
\tilde{P}_{s,h}^{\text{act}, d^{\text{p}}} = \frac{P_{s,h}^{\text{act}, d^{\text{p}}}}{\max(P_{s,h}^{\text{act}, d^{\text{p}}})}
\end{equation}
here, $\max(\cdot)$ denotes the maximum value of the time series over all durations $t^{\text{p}} \in \Omega$, seasons $s \in \mathcal{S}$, and hours $h \in \mathcal{H}$.
Since the decision variable $t^{\text{p}}$ is restricted to a finite set of discrete values $\Omega$, the optimization problem reduces to selecting the optimal value of $t^{\text{p}}$. This can be achieved by simulating all values of $t^{\text{p}}$ and exploring the trade-off of the two objectives.

\section{Case study}
\label{sec:cstudy}
The proposed approach is tested on a VPP relative to a low-voltage network located in northern Switzerland, synthetically generated in~\cite{Oneto_2023}. The network comprises 97 buses and 96 lines and has \SI{150}{kW} of peak non-dispatchable load. DERs and load profiles are allocated using a GIS-based approach based on geographically referenced open-source data. DER penetration reflects projected 2030 levels~\cite{EnergyPerspectives2050+}, resulting in \SI{150}{kWp} of rooftop PV, \SI{85}{kW} of installed heat pumps, \SI{75}{kWh} of BESS, and 67 electric vehicles. EVs have an average battery capacity of \SI{70}{kWh} and a maximum charging power of \SI{7}{kW}. 

The analysis focuses on a power reserve booking product traded a day ahead. The product has a ramp rate requirement of $r^{\text{p}} = \SI{5}{min}$ and reliability requirement $R^{\text{p}} = \SI{99.9}{\%}$~\cite{swissgrid2022_prequalification}. DER profiles align with the representative days. The considered durations in hours are the integer divisors of 24, i.e. $\Omega = \{1,2,3,4,6,8,12,24\}$. Imbalance and DER profiles were computed for the year $\mathcal{Y} = \{2024\}$.

Uncertainty arises from day-ahead forecasting errors in solar irradiance, ambient temperature, electricity demand, and EV schedules. For electricity demand, solar irradiance, and ambient temperature, forecasting errors are assumed to follow a normal distribution with zero mean~\cite{6520086, VANDERMEER20181484} and thus reflect unbiased estimators. The standard deviations are derived from a literature review of state-of-the-art forecasting tools. They are set to \SI{8.15}{\%} for solar irradiance~\cite{solcast_forecast_accuracy}, \SI{1.50}{K} for ambient temperature~\cite{meteoblue_weather_accuracy}, and \SI{10.75}{\%} for demand~\cite{8039509}. Uncertainty in EV schedules is modeled using a different approach. The schedule from a mobility database~\cite{10694497} provides the upper bound (reference) for movements in the area. A schedule disruption parameter is introduced. A zero value implies that the realized schedule matches the database. Values between zero and one indicate that a randomly selected proportion of scheduled events equal to the disruption parameter is removed. This parameter is assumed to follow a uniform distribution from $\SI{0}{}$ to $\SI{20}{\%}$.

\begin{figure}
    \centering
    \includegraphics[width=\linewidth]{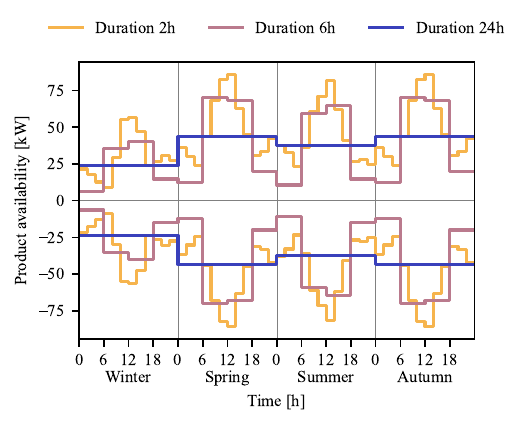}
    \caption{Reserve capacity supply profiles for different product durations. Positive values denote upward reserve and negative values downward reserve.}
    \label{fig:supply_curves}
\end{figure}

\section{Results and discussion}
\label{sec:results}
This section presents the results of applying the method described in Section~\ref{sec:method} on the case study VPP introduced in Section~\ref{sec:cstudy}. First, Section~\ref{sec:results_capacity_supply} presents the supply profiles and discusses their dependence on the product's duration. 
Then, Section~\ref{sec:results_product_design} presents the results of the product design.

\subsection{VPP Reserve capacity supply}
\label{sec:results_capacity_supply}

The reserve capacity supply profiles for the representative days were obtained using the methodology presented in Section~\ref{sec:method_reserve_supply} and reported in Figure~\ref{fig:supply_curves}. For ease of representation, only a subset of the product durations $\Omega$ is reported. However, the observations can be extended to intermediate product durations.
The profile data shows that the power reserve products are predominantly available during the daytime, especially in summer, spring, and autumn. This trend arises because the VPP's flexibility is largely provided by PV units, which are unavailable at night and less available during winter.

Furthermore, the impact of product duration on the supply profile is significant. As the product duration increases, the availability of reserve capacity during the day tends to decrease while the nighttime availability increases, reflecting an averaging effect. The choice of product duration strongly influences the shape of the supply profile. For instance, the 2-hour product exhibits higher peaks around noon than the 6-hour product, but is less available during PV ramp times. Conversely, the 24-hour product shows the highest availability during nighttime.
Another notable observation is the similarity between upward and downward reserve profiles. This can be attributed to the assumption that the ramping capabilities of the DERs in the VPP are symmetric in upward and downward directions. 

\subsection{Product design}
\label{sec:results_product_design}

\begin{figure}
    \centering
    \includegraphics[width=\linewidth]{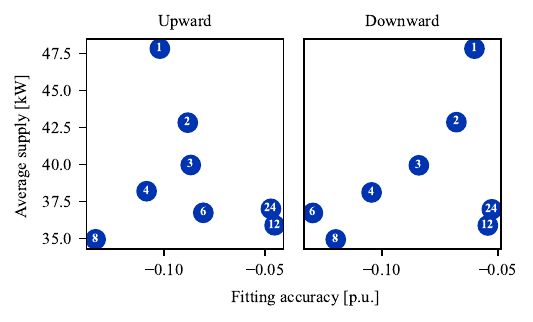}
    \caption{Value of the two objectives for each considered $t^{\text{p}} \in \Omega$, highlighted through the numbers in the data points. The left panel shows the upward reserve and the right panel shows the downward reserve. Better designs are located in the upper-right part of each graph.}
    \label{fig:pareto_scatter}
\end{figure}

\begin{figure}
    \centering
    \includegraphics[width=\linewidth]{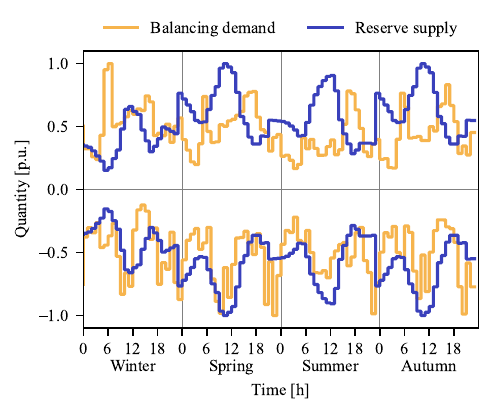}
    \caption{Balancing needs and reserve capacity supply profiles for 1-hour product duration. Positive values denote upward reserve and negative values downward reserve.}
    \label{fig:fitting_profiles}
\end{figure}

The reserve capacity supply profiles for each $t^{\text{p}} \in \Omega$ were processed as described in Section~\ref{sec:method_product_design} to compute the values of the two objectives, namely, average availability and alignment with system balancing needs. The results are illustrated in Figure~\ref{fig:pareto_scatter}. Both objectives aim to be maximized; therefore, better designs are located in the upper-right part of each graph.
Figure~\ref{fig:pareto_scatter} shows that short products (shorter than three hours) have higher average availability than longer ones. This trend highlights the advantage of short-duration products in exploiting flexibility during periods of high availability.
Conversely, longer product durations (e.g., 12 and 24 hours) achieve superior performance in terms of alignment with balancing needs. This can be attributed to the balancing demand profiles shown in Figure~\ref{fig:fitting_profiles}, which show periods of high demand during morning and evening PV ramps and night-time. Long-duration products are more capable of capturing these night-time peaks, thereby improving the alignment with the system's balancing needs.

We also observe that short products fit better downward reserve demand than upward products, resulting in larger values for the second objective. This can be explained by examining Figure~\ref{fig:fitting_profiles}, where downward balancing needs exhibit a pronounced day peak, which is well-matched by the availability profiles of short-duration products. In contrast, upward balancing needs display more complex patterns, with morning and evening peaks less effectively captured by the shorter durations.

All considered, the results highlight a trade-off between the two objectives. Choosing longer product durations enhances adherence of the supply profile to the balancing needs but reduces the available quantities. Conversely, shorter durations increase the average quantities but diminish the fit to the demand, particularly for upward reserve.

For downward reserve, selecting a short product duration sacrifices only a small degree of fitting accuracy while significantly increasing the available quantity. Among the durations in $\Omega$, the 1-hour product emerges as the optimal choice for downward reserve. For upward reserve, however, the trade-off is more pronounced. A 1-hour product maximizes availability, while 12-hour or 24-hour products better match the demand profile. To maintain consistency, a 1-hour duration could be adopted for upward reserve as well, despite the resulting upward reserve supply potential being less aligned with the balancing needs. 

In conclusion, we identify the 1-hour product as the best design, maximizing product availability and providing fair adherence to the balancing needs. However, being most available during the daytime, it leaves uncovered the morning and evening peaks for upward reserve and the night peak for downward reserve. Other service providers are required to address these gaps.



\section{Conclusion}
\label{sec:conclusions}
This paper proposes a methodology to design the optimal duration of reserve capacity ancillary service products for distributed energy resources. The method involves characterizing imbalance profiles from historical secondary reserve energy activation data. Then, the reserve capacity supply potential profiles of a DER aggregate are found, and they are parametrized according to the product duration. In this step, DER operational constraints, network limits, and product technical specifications under forecast uncertainty are considered. Finally, the optimal product duration is determined by maximizing product availability and aligning the supply profile with the system imbalance profile.

The proposed methodology is applied to a realistic Swiss low-voltage network with a diverse DER portfolio. The results demonstrate that shorter product durations (lower than three hours) maximize product availability, while longer durations (e.g., 12 or 24 hours) align best with system balancing needs. We find that a 1-hour product duration maximizes reserve availability while providing reasonable alignment with balancing needs. However, other service providers are needed to address demand peaks during morning and evening hours for upward reserve and night peaks for downward reserve.

In future work, we will extend the analysis to a wider range of DER portfolios to derive more generalizable design guidelines. Furthermore, the design of other AS product parameters (i.e., reliability, ramp rate, and lead time) will be investigated. Additionally, we plan to perform long-term forecasts of imbalance profile shapes and apply the design method to these forecasts. This will allow us to design product requirements that align with future balancing needs.

\section*{Acknowledgment}
\label{Acknowledgement}
The authors would like to thank Dr. Raphael Wu from Swissgrid AG for the support and insightful discussions.

\bibliographystyle{IEEEtran}
\bibliography{main}

\end{document}